
%
%
%
%
%
\nonstopmode
\catcode`\@=11 
%
%
%
\font\seventeenrm=cmr17

\font\twelverm=cmr12
\font\ninerm=cmr9
\font\sixrm=cmr6

\font\seventeenbf=cmbx12 at 17pt
\font\fourteenbf=cmbx12 at 14pt
\font\twelvebf=cmbx12
\font\ninebf=cmbx9
\font\sixbf=cmbx6

\font\seventeeni=cmmi12 at 17pt             \skewchar\seventeeni='177
\font\fourteeni=cmmi12 at 14pt              \skewchar\fourteeni='177
\font\twelvei=cmmi12                        \skewchar\twelvei='177
\font\ninei=cmmi9                           \skewchar\ninei='177
\font\sixi=cmmi6                            \skewchar\sixi='177

\font\seventeensy=cmsy10 scaled\magstep3    \skewchar\seventeensy='60
\font\fourteensy=cmsy10 scaled\magstep2     \skewchar\fourteensy='60
\font\twelvesy=cmsy10 at 12pt               \skewchar\twelvesy='60
\font\ninesy=cmsy9                          \skewchar\ninesy='60
\font\sixsy=cmsy6                           \skewchar\sixsy='60

\font\seventeenex=cmex10 scaled\magstep3
\font\fourteenex=cmex10 scaled\magstep2
\font\twelveex=cmex10 at 12pt

\font\ninex=cmex10 at 9pt
\font\sevenex=cmex10 at 9pt
\font\sixex=cmex10 at 6pt
\font\fivex=cmex10 at 5pt

\font\seventeensl=cmsl10 scaled\magstep3
\font\fourteensl=cmsl10 scaled\magstep2
\font\twelvesl=cmsl10 scaled\magstep1
\font\ninesl=cmsl10 at 9pt
\font\sevensl=cmsl10 at 7pt
\font\sixsl=cmsl10 at 6pt
\font\fivesl=cmsl10 at 5pt

\font\seventeenit=cmti12 scaled\magstep2
\font\fourteenit=cmti12 scaled\magstep1
\font\twelveit=cmti12

\font\seventeentt=cmtt12 scaled\magstep2
\font\fourteentt=cmtt12 scaled\magstep1
\font\twelvett=cmtt12

\font\seventeencp=cmcsc10 scaled\magstep3
\font\fourteencp=cmcsc10 scaled\magstep2
\font\twelvecp=cmcsc10 scaled\magstep1
\font\tencp=cmcsc10

\newfam\cpfam

\font\seventeenss=cmss17
\font\fourteenss=cmss12 at 14pt
\font\twelvess=cmss12
\font\tenss=cmss10
\font\niness=cmss9

\font\sevenss=cmss8 at 7pt
\font\sixss=cmss8 at 6pt
\font\fivess=cmss8 at 5pt
\newfam\ssfam
\newdimen\b@gheight             \b@gheight=12pt
\newcount\f@ntkey               \f@ntkey=0
\def\f@m{\afterassignment\samef@nt\f@ntkey=}
\def\samef@nt{\fam=\f@ntkey \the\textfont\f@ntkey\relax}
\def\rm{\f@m0 }
\def\mit{\f@m1 }         
\def\cal{\f@m2 }
\def\it{\f@m\itfam}
\def\sl{\f@m\slfam}
\def\bf{\f@m\bffam}
\def\tt{\f@m\ttfam}
\def\ssf{\f@m\ssfam}
\def\caps{\f@m\cpfam}
\def\seventeenpoint{\relax
    \textfont0=\seventeenrm          \scriptfont0=\twelverm
      \scriptscriptfont0=\ninerm
    \textfont1=\seventeeni           \scriptfont1=\twelvei
      \scriptscriptfont1=\ninei
    \textfont2=\seventeensy          \scriptfont2=\twelvesy
      \scriptscriptfont2=\ninesy
    \textfont3=\seventeenex          \scriptfont3=\twelveex
      \scriptscriptfont3=\ninex
    \textfont\itfam=\seventeenit    
    \textfont\slfam=\seventeensl    
      \scriptscriptfont\slfam=\ninesl
    \textfont\bffam=\seventeenbf     \scriptfont\bffam=\twelvebf
      \scriptscriptfont\bffam=\ninebf
    \textfont\ttfam=\seventeentt
    \textfont\cpfam=\seventeencp
    \textfont\ssfam=\seventeenss     \scriptfont\ssfam=\twelvess
      \scriptscriptfont\ssfam=\niness
    \samef@nt
    \b@gheight=17pt
    \setbox\strutbox=\hbox{\vrule height 0.85\b@gheight
                                depth 0.35\b@gheight width\z@ }}
\def\fourteenpoint{\relax
    \textfont0=\fourteencp          \scriptfont0=\tenrm
      \scriptscriptfont0=\sevenrm
    \textfont1=\fourteeni           \scriptfont1=\teni
      \scriptscriptfont1=\seveni
    \textfont2=\fourteensy          \scriptfont2=\tensy
      \scriptscriptfont2=\sevensy
    \textfont3=\fourteenex          \scriptfont3=\twelveex
      \scriptscriptfont3=\tenex
    \textfont\itfam=\fourteenit     \scriptfont\itfam=\tenit
    \textfont\slfam=\fourteensl     \scriptfont\slfam=\tensl
      \scriptscriptfont\slfam=\sevensl
    \textfont\bffam=\fourteenbf     \scriptfont\bffam=\tenbf
      \scriptscriptfont\bffam=\sevenbf
    \textfont\ttfam=\fourteentt
    \textfont\cpfam=\fourteencp
    \textfont\ssfam=\fourteenss     \scriptfont\ssfam=\tenss
      \scriptscriptfont\ssfam=\sevenss
    \samef@nt
    \b@gheight=14pt
    \setbox\strutbox=\hbox{\vrule height 0.85\b@gheight
                                depth 0.35\b@gheight width\z@ }}
\def\twelvepoint{\relax
    \textfont0=\twelverm          \scriptfont0=\ninerm
      \scriptscriptfont0=\sixrm
    \textfont1=\twelvei           \scriptfont1=\ninei
      \scriptscriptfont1=\sixi
    \textfont2=\twelvesy           \scriptfont2=\ninesy
      \scriptscriptfont2=\sixsy
    \textfont3=\twelveex          \scriptfont3=\ninex
      \scriptscriptfont3=\sixex
    \textfont\itfam=\twelveit    
    \textfont\slfam=\twelvesl    
      \scriptscriptfont\slfam=\sixsl
    \textfont\bffam=\twelvebf     \scriptfont\bffam=\ninebf
      \scriptscriptfont\bffam=\sixbf
    \textfont\ttfam=\twelvett
    \textfont\cpfam=\twelvecp
    \textfont\ssfam=\twelvess     \scriptfont\ssfam=\niness
      \scriptscriptfont\ssfam=\sixss
    \samef@nt
    \b@gheight=12pt
    \setbox\strutbox=\hbox{\vrule height 0.85\b@gheight
                                depth 0.35\b@gheight width\z@ }}
\def\tenpoint{\relax
    \textfont0=\tenrm          \scriptfont0=\sevenrm
      \scriptscriptfont0=\fiverm
    \textfont1=\teni           \scriptfont1=\seveni
      \scriptscriptfont1=\fivei
    \textfont2=\tensy          \scriptfont2=\sevensy
      \scriptscriptfont2=\fivesy
    \textfont3=\tenex          \scriptfont3=\sevenex
      \scriptscriptfont3=\fivex
    \textfont\itfam=\tenit     \scriptfont\itfam=\seveni
    \textfont\slfam=\tensl     \scriptfont\slfam=\sevensl
      \scriptscriptfont\slfam=\fivesl
    \textfont\bffam=\tenbf     \scriptfont\bffam=\sevenbf
      \scriptscriptfont\bffam=\fivebf
    \textfont\ttfam=\tentt
    \textfont\cpfam=\tencp
    \textfont\ssfam=\tenss     \scriptfont\ssfam=\sevenss
      \scriptscriptfont\ssfam=\fivess
    \samef@nt
    \b@gheight=10pt
    \setbox\strutbox=\hbox{\vrule height 0.85\b@gheight
                                depth 0.35\b@gheight width\z@ }}
%
%
%
\normalbaselineskip = 15pt plus 0.2pt minus 0.1pt 
\normallineskip = 1.5pt plus 0.1pt minus 0.1pt
\normallineskiplimit = 1.5pt
\newskip\normaldisplayskip
\normaldisplayskip = 15pt plus 5pt minus 10pt 
\newskip\normaldispshortskip
\normaldispshortskip = 6pt plus 5pt
\newskip\normalparskip
\normalparskip = 6pt plus 2pt minus 1pt
\newskip\skipregister
\skipregister = 5pt plus 2pt minus 1.5pt
\newif\ifsingl@    \newif\ifdoubl@
\newif\iftwelv@    \twelv@true
\def\singlespace{\singl@true\doubl@false\spaces@t}
\def\doublespace{\singl@false\doubl@true\spaces@t}
\def\normalspace{\singl@false\doubl@false\spaces@t}
\def\Tenpoint{\tenpoint\twelv@false\spaces@t}
\def\Twelvepoint{\twelvepoint\twelv@true\spaces@t}
\def\spaces@t{\relax
      \iftwelv@ \ifsingl@\subspaces@t3:4;\else\subspaces@t1:1;\fi
       \else \ifsingl@\subspaces@t3:5;\else\subspaces@t4:5;\fi \fi
      \ifdoubl@ \multiply\baselineskip by 5
         \divide\baselineskip by 4 \fi }
\def\subspaces@t#1:#2;{
      \baselineskip = \normalbaselineskip
      \multiply\baselineskip by #1 \divide\baselineskip by #2
      \lineskip = \normallineskip
      \multiply\lineskip by #1 \divide\lineskip by #2
      \lineskiplimit = \normallineskiplimit
      \multiply\lineskiplimit by #1 \divide\lineskiplimit by #2
      \parskip = \normalparskip
      \multiply\parskip by #1 \divide\parskip by #2
      \abovedisplayskip = \normaldisplayskip
      \multiply\abovedisplayskip by #1 \divide\abovedisplayskip by #2
      \belowdisplayskip = \abovedisplayskip
      \abovedisplayshortskip = \normaldispshortskip
      \multiply\abovedisplayshortskip by #1
        \divide\abovedisplayshortskip by #2
      \belowdisplayshortskip = \abovedisplayshortskip
      \advance\belowdisplayshortskip by \belowdisplayskip
      \divide\belowdisplayshortskip by 2
      \smallskipamount = \skipregister
      \multiply\smallskipamount by #1 \divide\smallskipamount by #2
      \medskipamount = \smallskipamount \multiply\medskipamount by 2
      \bigskipamount = \smallskipamount \multiply\bigskipamount by 4 }
\def\normalbaselines{ \baselineskip=\normalbaselineskip
   \lineskip=\normallineskip \lineskiplimit=\normallineskip
   \iftwelv@\else \multiply\baselineskip by 4 \divide\baselineskip by 5
     \multiply\lineskiplimit by 4 \divide\lineskiplimit by 5
     \multiply\lineskip by 4 \divide\lineskip by 5 \fi }
\Twelvepoint  
%
\interlinepenalty=50
\interfootnotelinepenalty=5000
\predisplaypenalty=9000
\postdisplaypenalty=500
\hfuzz=1pt
\vfuzz=0.2pt
\dimen\footins=24 truecm 
\hoffset=10.5truemm 
\voffset=-8.5 truemm 
%
%
%
%
%
%
\def\footnote#1{\edef\@sf{\spacefactor\the\spacefactor}#1\@sf
      \insert\footins\bgroup\singl@true\doubl@false\Tenpoint
      \interlinepenalty=\interfootnotelinepenalty \let\par=\endgraf
        \leftskip=\z@skip \rightskip=\z@skip
        \splittopskip=10pt plus 1pt minus 1pt \floatingpenalty=20000
        \smallskip\item{#1}\bgroup\strut\aftergroup\@foot\let\next}
\skip\footins=\bigskipamount 
\dimen\footins=24truecm 
\newcount\fnotenumber
\def\clearfnotenumber{\fnotenumber=0}
\def\fnote{\advance\fnotenumber by1 \footnote{$^{\the\fnotenumber}$}}
\clearfnotenumber
%
%
\newcount\secnumber
\newcount\appnumber
\newif\ifs@c 
\newif\ifs@cd 
\s@cdtrue 
\def\unsectioned{\s@cdfalse\let\section=\subsection}
\def\clearappnumber{\appnumber=64}
\def\clearsecnumber{\secnumber=0}
\newskip\sectionskip         \sectionskip=\medskipamount
\newskip\headskip            \headskip=8pt plus 3pt minus 3pt
\newdimen\sectionminspace    \sectionminspace=10pc
\newdimen\referenceminspace  \referenceminspace=25pc
\def\Titlestyle#1{\par\begingroup \interlinepenalty=9999
     \leftskip=0.02\hsize plus 0.23\hsize minus 0.02\hsize
     \rightskip=\leftskip \parfillskip=0pt
     \advance\baselineskip by 0.5\baselineskip
     \hyphenpenalty=9000 \exhyphenpenalty=9000
     \tolerance=9999 \pretolerance=9000
     \spaceskip=0.333em \xspaceskip=0.5em
     \seventeenpoint
  \noindent #1\par\endgroup }
\def\titlestyle#1{\par\begingroup \interlinepenalty=9999
     \leftskip=0.02\hsize plus 0.23\hsize minus 0.02\hsize
     \rightskip=\leftskip \parfillskip=0pt
     \hyphenpenalty=9000 \exhyphenpenalty=9000
     \tolerance=9999 \pretolerance=9000
     \spaceskip=0.333em \xspaceskip=0.5em
     \fourteenpoint
   \noindent #1\par\endgroup }
%
\def\spacecheck#1{\dimen@=\pagegoal\advance\dimen@ by -\pagetotal
   \ifdim\dimen@<#1 \ifdim\dimen@>0pt \vfil\break \fi\fi}
\def\section#1{\cleareqnumber \s@ctrue \global\advance\secnumber by1
   \message{Section \the\secnumber: #1}
   \par \ifnum\the\lastpenalty=30000\else
   \penalty-200\vskip\sectionskip \spacecheck\sectionminspace\fi
   \noindent {\caps\enspace\S\the\secnumber\quad #1}\par
   \nobreak\vskip\headskip \penalty 30000 }
\def\subsection#1{\par
   \ifnum\the\lastpenalty=30000\else \penalty-100\smallskip
   \spacecheck\sectionminspace\fi
   \noindent\undertext{#1}\enspace \vadjust{\penalty5000}}

\def\undertext#1{\vtop{\hbox{#1}\kern 1pt \hrule}}
\def\subsubsection#1{\par
   \ifnum\the\lastpenalty=30000\else \penalty-100\smallskip \fi
   \noindent\hbox{#1}\enspace \vadjust{\penalty5000}}

\def\appendix#1{\cleareqnumber \s@cfalse \global\advance\appnumber by1
   \message{Appendix \char\the\appnumber: #1}
   \par \ifnum\the\lastpenalty=30000\else
   \penalty-200\vskip\sectionskip \spacecheck\sectionminspace\fi
   \noindent {\caps\enspace Appendix \char\the\appnumber\quad #1}\par
   \nobreak\vskip\headskip \penalty 30000 }
\clearsecnumber
\clearappnumber
%
%
\def\ack{\par\penalty-100\medskip \spacecheck\sectionminspace
   \line{\iftwelv@\fourteencp\else\twelvecp\fi\hfil ACKNOWLEDGEMENTS\hfil}%
\nobreak\vskip\headskip }
\def\refs{\begingroup \par\penalty-100\medskip \spacecheck\sectionminspace
   \line{\iftwelv@\fourteencp\else\twelvecp\fi\hfil REFERENCES\hfil}%
\nobreak\vskip\headskip \frenchspacing }
\def\endrefs{\par\endgroup}
%
\newcount\refnumber
\def\clearrefnumber{\refnumber=0}  \clearrefnumber
\newwrite\R@fs                              
\immediate\openout\R@fs=\jobname.references 
\def\closerefs{\immediate\closeout\R@fs} 
\def\refsout{\closerefs\refs
\catcode`\@=11                          
\input\jobname.references               
\catcode`\@=12			        
\endrefs}
\def\refitem#1{\item{{\bf #1}}}
\def\ifundefined#1{\expandafter\ifx\csname#1\endcsname\relax}
%
%
\def\[#1]{\ifundefined{#1R@FNO}%
\global\advance\refnumber by1%
\expandafter\xdef\csname#1R@FNO\endcsname{[\the\refnumber]}%
\immediate\write\R@fs{\noexpand\refitem{\csname#1R@FNO\endcsname}%
\noexpand\csname#1R@F\endcsname}\fi{\bf \csname#1R@FNO\endcsname}}
\def\refdef[#1]#2{\expandafter\gdef\csname#1R@F\endcsname{{#2}}}
%
%
%
%
%
%
\newcount\eqnumber
\def\cleareqnumber{\eqnumber=0}
\newif\ifal@gn \al@gnfalse  
\def\veqnalign#1{\al@gntrue \vbox{\eqalignno{#1}} \al@gnfalse}
\def\eqnalign#1{\al@gntrue \eqalignno{#1} \al@gnfalse}
\def\(#1){\relax%
\ifundefined{#1@Q}
 \global\advance\eqnumber by1
 \ifs@cd
  \ifs@c
   \expandafter\xdef\csname#1@Q\endcsname{{%
\noexpand\rm(\the\secnumber .\the\eqnumber)}}
  \else
   \expandafter\xdef\csname#1@Q\endcsname{{%
\noexpand\rm(\char\the\appnumber .\the\eqnumber)}}
  \fi
 \else
  \expandafter\xdef\csname#1@Q\endcsname{{\noexpand\rm(\the\eqnumber)}}
 \fi
 \ifal@gn
    & \csname#1@Q\endcsname
 \else
    \eqno \csname#1@Q\endcsname
 \fi
\else%
\csname#1@Q\endcsname\fi\global\let\@Q=\relax}
%
%
\newif\iffrontpage \frontpagefalse
\headline={\hfil}
\footline={\iffrontpage\hfil\else \hss\twelverm
-- \folio\ --\hss \fi }
\def\monthname{\relax\ifcase\month 0/\or January\or February\or
   March\or April\or May\or June\or July\or August\or September\or
   October\or November\or December\else\number\month/\fi}
\hsize=14 truecm
\vsize=22 truecm
\skip\footins=\bigskipamount
\normalspace
%
%
%
\newskip\frontpageskip
\newif\ifp@bblock \p@bblocktrue
\newif\ifm@nth \m@nthtrue
\newtoks\pubnum
\newtoks\pubtype
\newtoks\m@nthn@me
\newcount\Ye@r
\advance\Ye@r by \year
\advance\Ye@r by -1900
\def\Year#1{\Ye@r=#1}
\def\Month#1{\m@nthfalse \m@nthn@me={#1}}
\def\m@nthname{\ifm@nth\monthname\else\the\m@nthn@me\fi}
\def\titlepage{\global\frontpagetrue\hrule height\z@ \relax
               \ifp@bblock\pubblock\fi\relax }
\def\endtitlepage{\vfil\break
                  \frontpagefalse} 
\def\bonntitlepage{\global\frontpagetrue\hrule height\z@ \relax
               \ifp@bblock\pubblock\fi\relax }
\frontpageskip=12pt plus .5fil minus 2pt
\pubtype={\iftwelv@\twelvesl\else\tensl\fi\ (Preliminary Version)}
\pubnum={?}
\def\nopubblock{\p@bblockfalse}
\def\pubblock{\line{\hfil\iftwelv@\twelverm\else\tenrm\fi%
BONN--HE--\number\Ye@r--\the\pubnum\the\pubtype}
              \line{\hfil\iftwelv@\twelverm\else\tenrm\fi%
\m@nthname\ \number\year}}
\def\title#1{\vskip\frontpageskip\Titlestyle{\caps #1}\vskip3\headskip}
\def\author#1{\vskip.5\frontpageskip\titlestyle{\caps #1}\nobreak}

\def\address#1{\par\kern 5pt\titlestyle{
\it #1}}
\def\andaddress{\par\kern 5pt \centerline{\sl and} \address}

\def\KUL{\address{Instituut voor Theoretische Fysica, Universiteit Leuven\break
Celestijnenlaan 200 D, B--3001 Heverlee, BELGIUM}}
\def\Santiago{\address{Departamento de F{\'\i}sica de Part{\'\i}culas
Elementales\break
Universidad de Santiago, Santiago de Compostela 15706, SPAIN}}
\def\abstract#1{\par\dimen@=\prevdepth \hrule height\z@ \prevdepth=\dimen@
   \vskip\frontpageskip\spacecheck\sectionminspace
   \centerline{\iftwelv@\fourteencp\else\twelvecp\fi ABSTRACT}\vskip\headskip
   {\noindent #1}}
%

%
%
%
\def\leaderfill{\leaders\hbox to 1em{\hss.\hss}\hfill}
\def\boxit#1{\vcenter{\hrule\hbox{\vrule\kern8pt
      \vbox{\kern8pt#1\kern8pt}\kern8pt\vrule}\hrule}}

%
%
%
\def\ref#1{{\bf [#1]}}
\def\ie{{\it i.e.\/}}
\def\nl{\hfil\break}
%
%
%
%
%
\newif\ifm@thstyle \m@thstylefalse
\def\mathstyle{\m@thstyletrue}
\def\proclaim#1#2\par{\smallbreak\begingroup
\advance\baselineskip by -0.25\baselineskip%
\advance\belowdisplayskip by -0.35\belowdisplayskip%
\advance\abovedisplayskip by -0.35\abovedisplayskip%
    \noindent{\caps#1.\enspace}{#2}\par\endgroup%
\smallbreak}
\def\m@kem@th<#1>#2#3{%
\ifm@thstyle \global\advance\eqnumber by1
 \ifs@cd
  \ifs@c
   \expandafter\xdef\csname#1\endcsname{{%
\noexpand #2\ \the\secnumber .\the\eqnumber}}
  \else
   \expandafter\xdef\csname#1\endcsname{{%
\noexpand #2\ \char\the\appnumber .\the\eqnumber}}
  \fi
 \else
  \expandafter\xdef\csname#1\endcsname{{\noexpand #2\ \the\eqnumber}}
 \fi
 \proclaim{\csname#1\endcsname}{#3}
\else
 \proclaim{#2}{#3}
\fi}
%
%
%
%
%
%
\def\Thm<#1>#2{\m@kem@th<#1M@TH>{Theorem}{\sl#2}}
\def\Prop<#1>#2{\m@kem@th<#1M@TH>{Proposition}{\sl#2}}
\def\Def<#1>#2{\m@kem@th<#1M@TH>{Definition}{\rm#2}}
\def\Lem<#1>#2{\m@kem@th<#1M@TH>{Lemma}{\sl#2}}
\def\Cor<#1>#2{\m@kem@th<#1M@TH>{Corollary}{\sl#2}}
\def\Conj<#1>#2{\m@kem@th<#1M@TH>{Conjecture}{\sl#2}}
\def\Rmk<#1>#2{\m@kem@th<#1M@TH>{Remark}{\rm#2}}
\def\Exm<#1>#2{\m@kem@th<#1M@TH>{Example}{\rm#2}}
\def\Qry<#1>#2{\m@kem@th<#1M@TH>{Query}{\it#2}}
%
%
\let\Pf=\Proof

\def\<#1>{\csname#1M@TH\endcsname}
%
%
\def\qed{\vrule width 0.7em height 0.6em depth 0.2em}

\def\iff{\Leftrightarrow}
\def\lapprox{\hbox{\lower3pt\hbox{$\buildrel<\over\sim$}}}
\def\gapprox{\hbox{\lower3pt\hbox{$\buildrel<\over\sim$}}}
\def\quotient#1#2{#1/\lower0pt\hbox{${#2}$}}
%
%
\def\to{\rightarrow}
%

%
\def\comap#1{\buildrel {#1} \over\longrightarrow} 
%
%
\def\reals{{\bf R}} 
\def\integ{{\bf Z}} 
\def\nats{{\bf N}} 
%
%
\def\Tr{{\rm Tr}}
\def\underrightarrow#1{\vtop{\ialign{##\crcr
      $\hfil\displaystyle{#1}\hfil$\crcr
      \noalign{\kern-\p@\nointerlineskip}
      \rightarrowfill\crcr}}} 
\def\underleftarrow#1{\vtop{\ialign{##\crcr
      $\hfil\displaystyle{#1}\hfil$\crcr
      \noalign{\kern-\p@\nointerlineskip}
      \leftarrowfill\crcr}}}  

%
%
\def\comm#1#2{\left[#1\, ,\,#2\right]}
%
%
\def\pder#1#2{{{\partial #1}\over{\partial #2}}}
\def\dpder#1#2#3{{{\partial^2#1}\over{{\partial#2}{\partial#3}}}}%
\def\der#1#2{{{d #1}\over {d #2}}}
%
%
%
%
%
\newdimen\unit
\newdimen\redunit
%
%
\def\p@int#1:#2 #3 {\rlap{\kern#2\unit
     \raise#3\unit\hbox{#1}}}
%
%
\def\th@r{\vrule height0\unit depth.1\unit width1\unit}
\def\bh@r{\vrule height.1\unit depth0\unit width1\unit}
\def\lv@r{\vrule height1\unit depth0\unit width.1\unit}
\def\rv@r{\vrule height1\unit depth0\unit width.1\unit}
%
%
\def\t@ble@u{\hbox{\p@int\bh@r:0 0
                   \p@int\lv@r:0 0
                   \p@int\rv@r:.9 0
                   \p@int\th@r:0 1
                   }
             }
%
%
\def\t@bleau#1#2{\rlap{\kern#1\redunit
     \raise#2\redunit\t@ble@u}}
%
%
\newcount\n
\newcount\m
\def\makecol#1#2#3{\n=0 \m=#3
  \loop\ifnum\n<#1{}\advance\m by -1 \t@bleau{#2}{\number\m}\advance\n by 1
\repeat}
%
%
\def\makerow#1#2#3{\n=0 \m=#3
 \loop\ifnum\n<#1{}\advance\m by 1 \t@bleau{\number\m}{#2}\advance\n by 1
\repeat}
%
%
\def\checkunits{\ifinner \unit=6pt \else \unit=8pt \fi
                \redunit=0.9\unit } 
\def\ytsym#1{\checkunits\kern-.5\unit
  \vcenter{\hbox{\makerow{#1}{0}{0}\kern#1\unit}}\kern.5em} 
\def\ytant#1{\checkunits\kern.5em
  \vcenter{\hbox{\makecol{#1}{0}{0}\kern1\unit}}\kern.5em} 
\def\ytwo#1#2{\checkunits
  \vcenter{\hbox{\makecol{#1}{0}{0}\makecol{#2}{1}{0}\kern2\unit}}
                  \ } 
\def\ythree#1#2#3{\checkunits
  \vcenter{\hbox{\makecol{#1}{0}{0}\makecol{#2}{1}{0}\makecol{#3}{2}{0}%
\kern3\unit}}
                  \ } 
%
%
%

\def\NPB#1#2#3{{\sl Nucl. Phys.} {\bf B#1} (#2) #3}

\def\CMP#1#2#3{{\sl Comm. Math. Phys.} {\bf #1} (#2) #3}

\def\PLA#1#2#3{{\sl Phys. Lett.} {\bf #1A} (#2) #3}
\def\PLB#1#2#3{{\sl Phys. Lett.} {\bf #1B} (#2) #3}
\def\JMP#1#2#3{{\sl J. Math. Phys.} {\bf #1} (#2) #3}

\def\AoP#1#2#3{{\sl Ann. of Phys.} {\bf #1} (#2) #3}

\def\FAP#1#2#3{{\sl Funkt. Anal. Prilozheniya} {\bf #1} (#2) #3}
\def\FAaIA#1#2#3{{\sl Functional Analysis and Its Application} {\bf #1} (#2)
#3}

\def\Invm#1#2#3{{\sl Invent. math.} {\bf #1} (#2) #3}

\def\IJMPA#1#2#3{{\sl Int. J. Mod. Phys.} {\bf A#1} (#2) #3}
\def\AdM#1#2#3{{\sl Advances in Math.} {\bf #1} (#2) #3}

\def\TMP#1#2#3{{\sl Theor. Mat. Phys.} {\bf #1} (#2) #3}

\def\JSM#1#2#3{{\sl J. Soviet Math.} {\bf #1} (#2) #3}

\def\PJAS#1#2#3{{\sl Proc. Jpn. Acad. Sci.} {\bf #1} (#2) #3}
\def\JPSJ#1#2#3{{\sl J. Phys. Soc. Jpn.} {\bf #1} (#2) #3}
\def\JETPL#1#2#3{{\sl  Sov. Phys. JETP Lett.} {\bf #1} (#2) #3}
\catcode`\@=12 
%
%
%
%
\def\d{\partial}
\let\pb=\anticomm
\def\pdo{{\hbox{$\Psi$DO}}}
\def\fr#1/#2{\hbox{${#1}\over{#2}$}}
\def\Fr#1/#2{{{#1}\over{#2}}}

\def\Tr{{\rm Tr\,}}

\def\dlb#1#2{\lbrack\!\lbrack#1,#2\rbrack\!\rbrack}

\def\ope#1#2{{{#2}\over{\ifnum#1=1 {x-y} \else {(x-y)^{#1}}\fi}}}

\def\Gr{{\rm Gr}}
\def\gr{{\rm gr}}

\def\L{{\cal L}}
\refdef[Adler]{M.~Adler, \Invm{50}{1981}{403}.}
\refdef[MaRa]{Yu.~I.~Manin and A.~O.~Radul, \CMP{98}{1985}{65}.}
\refdef[DickeyI]{L.~A.~Dickey, {\sl Lectures in field theoretical
Lagrange-Hamiltonian formalism}, (unpublished).}
\refdef[DickeyII]{L.~A.~Dickey, \CMP{87}{1982}{127}.}
\refdef[DickeyIII]{L.~A.~Dickey, {\sl Integrable equations and Hamiltonian
systems}, World Scientific Publ.~Co.}
\refdef[GD]{I.~M.~Gel'fand and L.~A.~Dickey, {\sl A family of Hamiltonian
structures connected with integrable nonlinear differential equations},
Preprint 136, IPM AN SSSR, Moscow (1978).}
\refdef[DS]{V.~G.~Drinfel'd and V.~V.~Sokolov, \JSM{30}{1984}{1975}.}
\refdef[KW]{B.~A.~Kupershmidt and G.~Wilson, \Invm{62}{1981}{403}.}
\refdef[Mag]{F.~Magri, \JMP{19}{1978}{1156}.}
\refdef[Dickeypc]{L.~A.~Dickey, private communication.}
\refdef[STS]{M.~A.~Semenov-Tyan-Shanski\u\i, \FAaIA{17}{1983}{259}.}
\refdef[LM]{D.~R.~Lebedev and Yu.~I.~Manin, \FAP{13}{1979}{40}.}
\refdef[Uniw]{J.~M.~Figueroa-O'Farrill and E.~Ramos, {\sl Existence
and Uniqueness of the Universal $W$-Algebra}, to appear in
the {\sl J.~Math.~Phys.}}
\refdef[FL]{V.~A.~Fateev and S.~L.~Lykyanov, \IJMPA{3}{1988}{507}.}
\refdef[DFIZ]{P.~Di Francesco, C.~Itzykson, and J.-B.~Zuber,
\CMP{140}{1991}{543}.}
\refdef[KR]{V.~G.~Ka{\v c} and A.~C.~Raina, {\sl Lectures on highest
weight representations of infinite dimensional Lie algebras}, World
Scientific, etc...}
\refdef[KP]{E.~Date, M.~Jimbo, M.~Kashiwara, and T.~Miwa
\PJAS{57A}{1981}{387}; \JPSJ{50}{1981}{3866}.}
\refdef[DickeyKP]{L.~A.~Dickey, {\sl Annals of the New York Academy of
Science} {\bf 491} (1987) 131.}
\refdef[WKP]{J.~M.~Figueroa-O'Farrill, J.~Mas, and E.~Ramos,
\PLB{266}{1991}{298}.}
\refdef[Yama]{K. Yamagishi, \PLB{259}{1991}{436}.}
\refdef[YuWu]{F. Yu and Y.-S. Wu, {\sl Hamiltonian Structure,
(Anti-)Self-Adjoint Flows in KP Hierarchy and the $W_{1+\infty}$ and
$W_\infty$ Algebras}, Utah Preprint, January 1991.}
\refdef[ClassLim]{J. M. Figueroa-O'Farrill and E. Ramos, {\sl The
Classical Limit of $W\!$-Algebras}, Preprint-KUL-TF-92/5
and BONN-HE-92-03, February 1992.}
\refdef[KoSt]{B. Kostant and S. Sternberg, \AoP{176}{1987}{49}.}
\refdef[TakaTake]{K. Takasaki and T. Takebe, {\sl SDIFF(2) KP
Hierarchy}, Preprint RIMS-814, December 1991.}
\refdef[KoGi]{Y. Kodama, \PLA{129}{1988}{223},
\PLA{147}{1990}{477};\nl
Y. Kodama and J. Gibbons, \PLA{135}{1989}{167}.}
\refdef[Bakas]{I. Bakas, \CMP{134}{1990}{487}.}
\refdef[Pope]{C. N. Pope, L. J. Romans, and X. Shen,
\PLB{242}{1990}{401}.}
\refdef[Lang]{S. Lang, {\sl Algebra}, Second Edition, Addison-Wesley
1984.}
\refdef[Radul]{A. O. Radul, \JETPL{50}{1989}{373}.}
\refdef[Morozov]{A. Morozov, \NPB{357}{1991}{619}.}
\refdef[LPSW]{H. Lu, C. N. Pope, X. Shen, and X. J. Wang, {\sl The
complete structure of $W_N$ from $W_\infty$ at $c=-2$}, Texas A\& M
Preprint CPT TAMU-33/91 (May 1991).}
\refdef[Winfty]{C. N. Pope, L. J. Romans, and X. Shen,
\PLB{236}{1990}{173},\PLB{242}{1990}{401},\NPB{339}{1990}{191};\nl
I. Bakas, \CMP{134}{1990}{487}.}
\refdef[Zam]{A. B. Zamolodchikov, \TMP{65}{1986}{1205}.}
\refdef[Douglas]{M. R. Douglas, \PLB{238}{1990}{176}.}
\refdef[Guill]{V. W. Guillemin, \AdM{10}{1985}{131}.}
\refdef[Wod]{M.~Wodzicki, {\sl Noncommutative Residue}, in {\sl K-Theory,
Arithmetic and Geometry.} Ed. Yu. I. Manin. Lectures Notes in
Mathematics 1289, Springer-Verlag.}
\refdef[Shubin]{M.A. Shubin, {\sl Pseudodifferential operators
and spectral Theory}, Springer-Verlag.}
\refdef[Class]{J.~M.~Figueroa-O'Farrill and E.~Ramos, {\sl
Classical $W$-algebras from dispersionless Lax hierarchies
}, Preprint-KUL-TF-92/6, June 1992.}
\mathstyle
\overfullrule=0pt
\nopubblock
\titlepage
\line{\hfil Preprint-KUL-TF-92/19}
\line{\hfil US-FT/6-92}
\line{\hfil June 1992}
\title{Higher dimensional classical $W$-algebras}
\vskip 1.cm
\centerline{\it by}
\author{Fernando Martinez-Moras\footnote{$^\natural$}{\tt
e-mail: Fernando@gaes.usc.es\hfil}}
\Santiago
\vskip 0.5cm
\centerline{\it and}
\author{Eduardo Ramos\footnote{$^\sharp$}{\tt
e-mail: fgbda06@blekul11.BITNET.}\hfil\break
Address after October 1992:
Queen Mary and Westfield College,\hfil\break
Univ.~of London, U.K.\hfil}
\KUL

\abstract{Classical $W$-algebras in higher dimensions are constructed.
This is achieved by generalizing the classical Gel'fand-Dickey
brackets to the commutative limit of the ring of
classical pseudodifferential
operators in arbitrary dimension. These $W$-algebras are the Poisson
structures associated with a higher dimensional version of the
Khokhlov-Zabolotskaya hierarchy (dispersionless KP-hierarchy).
The two dimensional case is worked out explicitly and it is
shown that the role of Diff$S(1)$ is taken by the algebra of
generators of local diffeomorphisms in two dimensions.
}

\endtitlepage
\section{Introduction.}

$W$-algebras play a prominent role in two dimensional physics.
They first appeared in the context of integrable models
(although under a different name) as Poisson structures associated
with generalized KdV hierarchies$^{\[GD] ,\[Adler] ,\[DickeyIII]}$, but their
``popularity'' dramatically increased after the work of Zamolodchikov.
He showed in \[Zam], using the bootstrap method, that the simplest
extension of the Virasoro algebra by a field of spin 3 required the
introduction of a nonlinear associative algebra, denoted since then
by $W_3$. Soon after, Fateev and Lukyanov$^{\[FL]}$, using the formalism
developed by Drinfel'd and Sokolov$^{\[DS]}$, which relates to each generalized
KdV hierarchy a loop algebra, were able to generalize the results of
Zamolodchikov to construct $W_n$-algebras, {\ie} conformally extended
algebras with fields of integer spins from 3 to $n$.

Before continuing any further, we should clarify some notational
issues. In what follows, we will use the name $W$-algebras for
the quantum algebras. These are the ones realized in a conformal
field theory via operators acting on a Hilbert space. The
Gel'fand-Dickey algebras and their reductions
will be considered classical realizations of $W$-algebras.
We will reserve the name classical (one-dimensional)
$W$-algebras for nonlinear extensions of Diff$S(1)$.

Recently an unexpected connection has been unveiled between
Gel'fand-Dickey algebras (and their associated integrable
hierarchies), 2-D gravity, and, through their matrix model
formulation, noncritical strings coupled to $c<1$ matter$^{\[Douglas]}$.
In particuar, it has been
shown in \[Class] that the planar limit of these theories, in
which only manifolds with the topology of a sphere are considered, is
directly related to the Khokhlov-Zabolotskaya (KZ)
hierarchy$^{\[TakaTake] ,\[KoGi]}$
and its reductions. This hierarchy is also known as the dispersionless
or classical KP-hierarchy, where the deformation parameter that takes
us from the dispersive to the nondispersive case
happens to be, in the context of noncritical
strings, nothing but the renormalized string coupling constant.
Moreover, the Poisson structure  associated with the
KZ-hierarchy ($w_{KP}$) was shown to be
the universal $W$-algebra associated with the
$w_n$-series, {\ie}  the classical limit of the $W_n$-algebras.
We believe that this makes classical $W$-algebras a
very interesting and fruitful field of study.

Nevertheless, as interesting as all this may be from the physical
point of view, for example, the theory of 2-D surfaces, domain walls in 3-D,
or critical phenomena in 2-D systems, from the point of view of a
particle physicist, all these developments are little more than
toy-models for the relevant higher dimensional case. It is the aim
of this work to generalize some of these structures to higher
dimensions. The hope is that they will be helpful
in understanding topics such as nonperturbative gravity in $D>2$ and
even noncritical strings coupled to $c>1$ matter.

We should also point out that the generalization of the Gel'fand-
Dickey formalism to higher dimension comes as a little surprise.
It has been repeatedly claimed in the literature that an essential
ingredient in the one dimensional case was the existence of an
invariant splitting compatible with the Adler trace, a property
that disappears when one moves to higher dimensions. We will
show, at least in the classical limit, that we can proceed without it.

The plan of the paper is as follows:

In Section 2, the required formalism to extend
the usual one dimensional results for classical $W$-algebras
to arbitrary dimension will be developed. This
generalization will be shown to be quite straightforward, although
it will require the introduction of some nonstandard machinery
such as Guillemin's symplectic trace$^{\[Guill],\[Wod]}$,
which we will discuss briefly.

In Section 3, we will revisit the one dimensional case. This is
important because the generalization to higher dimensions will
require the introduction of a splitting in the space of
pseudodifferential symbols which does not reduce to the standard
in one dimension. Nevertheless, we will prove that the induced
algebras are nothing but the standard $w_{\infty}$ and the
$n\rightarrow\infty$ limit of the $w_n$ series
\footnote\dag{These two algebras should not be confused.
This is nothing but another
example of the poor notational conventions that plague this field.
In fact, $w_{\infty}$ is nothing but a particular contraction of
$w_{n\rightarrow\infty}$.}.

In Section 4, we will explicitly display the algebras associated
with the space of classical pseudodifferential symbols in two
dimensions. We will show that the role of Diff$S(1)$ is taken by
the algebra of local diffeomorphisms in two dimensions. We expect
this connection with diffeomorphism algebras to extend to arbitrary
dimension although we have not yet proved it so.

In Section 5, we will show how these Poisson structures are associated
with a very natural generalization of the KZ-hierarchy in higher
dimensions. We will also briefly discuss their integrability.

Finally, in the conclusion we will recapitulate our results and also
comment about deformations of these
structures. We will explain why the standard deformation which
takes $w_n$ to $W_n$ in one dimension is not directly applicable to the
higher dimensional case.

\section{General formalism.}
The higher dimensional $W$-algebras are going to appear as Poisson
structures in the space of pseudodifferential operators (\pdo) of the
classical type, so before going any further we will introduce
the necessary concepts to deal with such objects$^{\[Shubin] ,\[Wod]}$.

A {\pdo} ${\cal B}$ is said to be of the classical type if
${\cal B} :C^{\infty}_{comp}(X)\rightarrow C^{\infty}(X)$
can be expressed locally as
$${\cal B}f(x)=\int_{\reals_{\xi}^D}\int_{\reals_y^D}
{\rm e}^{i(x-y)\xi}\Lambda (x,\xi)f(y) {d^D\xi\over (2\pi)^D}d^Dy
+ {\cal T}f(x),
\(PDO)$$
where $D$ is the dimension of $X$ and $\cal T$ is an operator with a smooth
kernel\footnote\dag{This roughly means that $\Lambda$ is uniquely
defined up to terms that decay faster than any power of $|\xi |$ when
$|\xi |\rightarrow\infty$}, and
$\Lambda (x, \xi)\in C^{\infty} (T^*X\backslash 0)$ ({\ie} smooth
functions on the cotangent bundle without the zero section) admits an
asymptotic expansion for $|\xi|\rightarrow\infty$ of the form
$$\Lambda =\sum_{j=0}^{\infty}\tilde u_{m-j},\(asymp)$$
with
$$\tilde u_{m-j}(x,t\xi)=t^{m-j}\tilde u_{m-j}(x,\xi)\ \ \  t>0\(hom)$$
which implies that
$$\tilde u_k(x,\xi)=u_k(x,\theta)|\xi |^k,\(redef)$$
where $\theta$ stands for the angular variables in $\reals^D$.

{}From now on when we refer to a {\pdo} or its (smoothed) symbol, the reader
should assume that it is of the classical type unless stated otherwise.

We will further restrict ourselves to \pdo 's with $u_m$ equal to a
constant, which we will take equal to one in order to simplify the
notation, and $m\in\integ$.
Then, $M^m_D$ will denote the space of formal series of
the form \(asymp) with the above constraint.

We can equip $M^m_D$ with a Lie algebra structure by defining the Lie
bracket for any two symbols $A$ and $B$ belonging to $M^m_D$ by
$$[A,B]=A\circ B -B\circ A\(bracket)$$
where $\circ$ denotes the usual composition of symbols given by
$$A\circ B=\sum_{\alpha}{1\over\alpha !}\partial^{\alpha}_{\xi}
A\partial_x^{\alpha}B\(composition)$$
where $\alpha$ is a multi-index, and a ``nasty''
factor of $-i$ has been absorbed in the definition of $\partial_x$.

Although, it would be very interesting to keep the full
noncommutative structure of this space (see coments about
deformations
in the conclusions),
in this paper we will restrict ourselves to the commutative limit,
which we define below.

\subsection{The Classical limit of the ring of Pseudodifferential
Symbols.}

Let $R^D$ denote the ring of pseudodifferential symbols in
dimension $D$, where
multiplication is defined by the composition of symbols.
It is possible
to define a degree in $R^D$ as follows. Let the degree of a monomial
be given by its degree of homogeneity,
$${\rm deg}\ \tilde u(x,\xi)=j\ \iff\ \tilde u(x,t\xi)=t^j
\tilde u(x,\xi)\()$$
for any $t>0$.
We can now define the degree of a polynomial as the degree of the
leading term. Let us denote by $R^D_p$ the subspace of all symbols with
degree equal to or smaller than $p$. It is clear from \(composition) that
$$R^D_p\circ R^D_q =R_{p+q}^D,\(filtration)$$
making $R^D$ into a filtered ring.
Starting from any filtered ring we can define its associated graded
ring as follows. Let
$$\Gr_p R^D\equiv R^D_p/R^D_{p-1}.\(gradation)$$

The multiplication in $R^D$ induces a multiplication in
$\Gr R^D\equiv\bigoplus_p\Gr_p R^D$. The reader can easily check that
the composition
$$R^D_p\times R^D_q\comap{\circ} R^D_{p+q}\comap{\gr _{p+q}}
\Gr_{p+q}R^D\(doulblemap)$$
induces a unique map
$$\Gr_p R^D\times\Gr_q R^D\rightarrow\Gr_{p+q} R^D,\(filtered ring)$$
converting $\Gr R^D$ into a graded ring. Moreover, this induced
multiplication is nothing but the usual commutative multiplication
of symbols as formal series. This would not be very interesting as it
stands were if not for the fact that $\Gr R^D$ can be given a natural
Poisson algebra structure.

The commutator $\comm{A}{B}$ in $R^D$ induces a Lie
bracket on $\Gr\, R^D$ as follows.  Let $A\in R^D_p$ and $B\in R^D_q$.  It
is then easy to see that $\comm{A}{B} \in R^D_{p+q-1}$. The part in
$R^D_{p+q}$ has to vanish since it corresponds to the commutator in
$\Gr\,R^D$, which is a commutative algebra.  Therefore $\comm{A}{B}$
defines an element in $\Gr_{p+q-1}R^D$.  Furthermore, this element only
depends on the class of $A$ modulo $R^D_{p-1}$, for if $A\in R^D_{p-1}$,
then $\comm{A}{B} \in R^D_{p+q-2}$.  Similarly, it only depends on the
class of $B$ modulo $R^D_{q-1}$.  Therefore the composition
$$R^D_p \times R^D_q \comap{\comm{}{}} R^D_{p+q-1} \comap{\gr_{p+q-1}}
\Gr_{p+q-1} R^D\()$$
induces a unique map
$$\pb{}{} : \Gr_p R^D \times \Gr_q R^D \to \Gr_{p+q-1} R^D~.\()$$
The fact that $\comm{}{}$ is a Lie bracket for $R^D$ means that
$\pb{}{}$ is a Lie bracket for $\Gr\,R^D$, but it has more structure.
It turns out that it is a derivation over the commutative
multiplication on $\Gr\,R^D$.  In fact, let $A\in R^D_p$, $B\in R^D_q$, and
$C\in R^D_s$.  Then
$$\veqnalign{&\pb{\gr_p(A)}{\gr_q(B) \gr_s(C)} =
\pb{\gr_p(A)}{\gr_{q+s}(BC)}\cr
&=\gr_{p+q+s-1} \comm{A}{BC}\cr
&=\gr_{p+q+s-1} (\comm{A}{B}C) + \gr_{p+q+s-1} (B\comm{A}{C})\cr
&=\gr_{p+q-1}\comm{A}{B} \gr_s(C) + \gr_q(B) \gr_{p+s-1}\comm{A}{C}\cr
&=\pb{\gr_p(A)}{\gr_q(B)} \gr_s(C) + \pb{\gr_p(A)}{\gr_s(C)}
\gr_q(B)~.\()\cr}$$
In summary, this turns $\Gr\,R^D$ into a Poisson algebra.
The reader will recognize that the Poisson bracket so defined is
nothing but the canonical Poisson bracket on a 2n-dimensional phase
space with canonical coordinates $(x^i,\xi_i)$.

We can also define this classical limit in a more ``physical'' way as
follows.  Let us introduce the formal parameter $\hbar$ in
\(composition)
$$A\circ_{\hbar} B\equiv\sum_{\alpha}{\hbar^{|\alpha |}\over
\alpha !}{\partial^{\alpha}A\over\partial\xi_{\alpha}}
{\partial^{\alpha}B\over\partial x^{\alpha}},\()$$
interpolating from the commutative multiplication for $\hbar =0$ to
the noncommutative composition of symbols for $\hbar =1$. For $\hbar$
different from zero we can reabsorb it by rescaling $\xi$. This
implies that $\circ_{\hbar}$ remains associative for all values of
$\hbar$.

The classical limit is given by the leading term in the
$\hbar\rightarrow 0$ limit. This implies that composition of symbols
goes to standard commutative multiplication. However, notice that for
the bracket, the leading term is already of order $\hbar$, therefore
$$\pb{A}{B}\equiv \lim_{\hbar\rightarrow 0}{1\over\hbar}
\comm{A}{B}_{\hbar}
=\sum_{j=1}^{D}\left(\pder{A}{\xi_j}\pder{B}{x^j}-
\pder{B}{\xi_j}\pder{A}{x^j}\right),\()$$
as we found before.

\subsection{Formal Geometry.}

The space $M^m_D$ can be given the structure of an infinite dimensional
manifold, but we will not need this machinery for our purposes. It
will be sufficient to endow $M^m_D$ with a formal geometry or
algebraization of the strictly necessary geometric concepts.
Our main goal is to define Poisson brackets on $M^m_D$. The
geometrical objects we should define are: the class of functions
on which we define the Poisson brackets, the vector fields, and
1-forms together with the map sending a function to its associated
hamiltonian vector field.

We will define Poisson brackets on functions of the form:
$$F[\Lambda ]=\int f(u),\()$$
where $f(u)$ is a polynomial of the $u_j$'s and
their derivatives. The precise meaning of the integration will
depend on the particular context, {\ie} what kind of functions
the $u$'s are or in which particular space they live.
In what follows, we will only
use the fact that $\int$ is a map which anhilates exact forms, {\ie}
there are no contributions from boundary terms.

The tangent space $T_{\Lambda}M^m_D$ at $\Lambda$ is isomorphic to
the infinitesimal deformations of $\Lambda$. These are clearly
pseudodifferential symbols belonging to $R^D_{m-1}$. If $A\in R^D_
{m-1}$ is of the form $A=\sum_{j\leq m-1}a_j(x,\theta)|\xi|^j$,
then the vector field $D_A$ acting on a function $F$ is defined
by
$$D_A F=\left.\der{}{\epsilon}\int f(u_j+\epsilon a_j) \right\vert_{
\epsilon=0}.\()$$

The cotangent space $T_{\Lambda}^{*}M^m_D$ will be defined as the dual
of $T_{\Lambda}M^m_D$ with respect to a nondegenerate inner product. The
required inner product is supplied by
Guillemin's symplectic trace$^{\[Guill] ,\[Wod]}$.

\subsection{Symplectic trace.}

Here we will adapt  the general
discussion of \[Wod] to our somewhat restricted interest.

Let $Y$ be a symplectic manifold of dimension 2D
and $\omega$ the corresponding
nondegenerate 2-form. Then the Poisson bracket for any two functions
in $Y$ is given by
$$\pb{f}{g}={\cal L}_{H_f}g=\omega (H_f ,H_g)=i(H_g)i(H_f)\omega ,\()$$
where
$$\omega(H_f,\cdot)=-df.\()$$

In our particular case $Y\equiv X\times\reals^D$ and $\omega$
is given locally by
$$\omega =\sum_{j=1}^D dx^j\wedge d\xi_j .\()$$

As we have already seen, there is a natural action of the
multiplicative group of the real numbers on $Y$ given by
$$\chi_t :\xi\mapsto t\xi\()$$
where $t\in\reals^{\times}_{+}$. This action is conformal, that is
$$\chi_t^*\omega =t\omega.\()$$

The infinitesimal action defines the Euler vector field
$$\sigma=\sum_{j=1}^D \xi_j\pder{}{\xi_j}.\()$$

Now we can define
$$\alpha =i(\sigma )\omega\ \ {\rm and}\ \  \mu =\alpha\wedge (d\alpha)
^{D-1}.\()$$

Notice that $d\alpha =di(\sigma)\omega=di(\sigma)\omega +
i(\sigma)d\omega={\cal L}_{\sigma}\omega =\omega$, where the
last equality is obvious because $\sigma$ is the vector field
generating the flow induced by $\chi_{exp\,t}$. This implies that
$\alpha$ is nothing but the canonical 1-form of classical mechanics.

It is now clear that
$$\chi_t^*\alpha =t\alpha\ \ {\rm and}\ \ \chi_t^*\mu =t^D\mu,\()$$
therefore
$${\cal L}_{\sigma}\alpha =\alpha,\ \ d\mu =\omega ^D,\ \
{\cal L}_{\sigma}\mu =D\mu.\()$$

We will now proceed to prove a couple of technical Lemmas which will be
required in what follows.

\Lem<I>{$$\pb{f}{g}\omega^D =Ddf\wedge dg\wedge\omega^{D-1}.$$}
\Pf
$$\eqalign{ 0=& i(H_f)i(H_g)\omega^{D+1}=-(D+1)i(H_f)(dg\wedge
\omega^D)\cr =& -(D+1)(i(H_f)dg\wedge\omega^D - dg\wedge i(H_f)
\omega^D)\cr =& -(D+1)(dg(H_f)\omega^D +Ddg\wedge df\wedge
\omega^{D-1})\cr =& (D+1)(Ddf\wedge dg\wedge\omega^{D-1}
-\pb{f}{g}\omega^D).\cr}$$
\line{\hfil \qed}

\Lem<II>{$$\pb{f}{g}\mu =d(gi(H_f)\mu )-(D-1)gdf\wedge\omega^{D-1}
-\L_{\sigma}(gdf)\wedge\omega^{D-1}.$$}
\Pf
$$\pb{f}{g}\mu ={1\over D}i(\sigma )\pb{f}{g}\omega^D .$$
Using the previous Lemma this can be written as
$$\eqalign{\pb{f}{g}\mu =& i(\sigma )(df\wedge dg\wedge\omega
^{D-1})\cr =& {1\over D}i(\sigma )(dg\wedge i(H_f)\omega^D)\cr
=& {1\over D}i(\sigma )d(gi(H_f)\omega^D),\cr}$$
where we have used $\L_{H_f}\omega =0$ in the last line. From this
we have,
$$\eqalign{\pb{f}{g}\mu =& {1\over D}\L_{\sigma}(gi(H_f)\omega^D)
-{1\over D}d(gi({\sigma})i(H_f)\omega^D)\cr
=& -\L_{\sigma}(gdf\wedge\omega^{D-1})+{1\over D}d(gi(H_f)
i(\sigma )\omega^D)\cr
=& -\L_{\sigma}(gdf)\wedge\omega^{D-1}-(D-1)gdf\wedge\omega^{D-1}
+d(gi(H_f)\mu ).\cr}$$
\line{\hfil \qed}

{}From now on we will only deal with homogeneous functions on $Y$.
If $f\in\Gr_p R^D$, then
$$\L_{\sigma} f =p f.\()$$

The following Lemma will prove to be fundamental.
\Lem<III>{If $f\in\Gr_p R^D$ and $g\in\Gr_q R^D$,
$$\pb{f}{g}\mu =d(gi(H_{f})\mu )-(p+q +D-1)g
df\wedge\omega
^{D-1}.$$}
\Pf It follows immediately from the two previous Lemmas and
$\L_{\sigma}(g df )=(p+q)g df$.\qed

Now we can define the symplectic residue by
$${\rm Res} f=\cases{f\mu &if $f\in\Gr_{-D}R^D$\cr
0& otherwise.\cr}\()$$

In order to define the symplectic trace notice that $Y$ is a
principal $\reals^{\times}_+$-bundle with base $Z$, {\ie} $Y/
\reals_+^{\times}= Z$.
The natural projection $Y\rightarrow Z$ will be denoted by
$\rho$. Notice that $f\mu$ is a $\reals_+^{\times}$ invariant
form (because $f\in\Gr_{-D}R^D\Rightarrow\L_{\sigma}(f\mu )=0$)
and since it is always horizontal ({\ie} annihilated by $i(\sigma )$),
a unique form $f\mu$ on $Z$ must exist such that
$$\rho^*f\mu =f\mu.$$

The symplectic Trace is then defined to be
$$\Tr f=\int_Z {\rm Res} f\(trace).$$

This definition is obviously independent on the section.

Now we can prove that the symplectic trace  defined above has the usual
trace property.

\Thm<Traza>{If $f\in\Gr_p R^D$ and $g\in\Gr_q R^D$, then $\Tr\pb{f}{g} =0$
for all $p$ and $q$.}

\Pf
$$\Tr \pb{f}{g}=\int_Z\pb{f}{g}\mu$$
if $\pb{f}{g}\in\Gr_{-D}R^D$. By \<III>
$$\Tr \pb{f}{g}=\int_Z d(gi(H_f)\mu
)-(p+q-1+D)gdf\wedge\omega^{D-1}=0.$$

The first term is zero because by assumption $\int$ annihilates
exact forms.
The second term also vanish by noticing that $\pb{f}{g}\in
\Gr_{p+q-1}R^D$.$\qed$

\subsection {Generalized Adler map.}

Now we have all the required ingredients to define Poisson brackets
on $M^m_D$. Remember that one of the crucial properties in the standard
one-dimensional classical case was that there were two closed
subspaces on $R^1$ under Poisson brackets$^{\[ClassLim]}$.
We will show that this is
also true in arbitrary dimensions if we modify the old
definitions slightly.

Define $R^D_{[q,p]}\equiv\bigoplus_{j=q}^p\Gr_j R^D$. It should be clear from
our previous discussion that $R^D_+\equiv R^D_{[1,\infty )}$ and
$R^D_-\equiv R^D_{(-\infty ,0]}$ are closed under Poisson brackets.
Notice that in one dimension this splitting differs from the standard
one.
It also will be convenient to define $R^D_{\oplus}\equiv R^D_{[-D,\infty )
}$ and $R^D_{\ominus}\equiv R^D_{(-\infty ,-D-1]}$.
For any {\pdo} symbol $A$, the symbols
$A_+$, $A_-$, $A_{\oplus}$ and $A_{\ominus}$
will denote the projections of $A$ on $R^D_+$, $R^D_-$, $R^D_{\oplus}$
and $R^D_{\ominus}$ respectively.

These $\oplus$ and $\ominus$ splittings will
appear naturally because, in contrast with the usual one, ours
is not compatible with the symplectic trace, {\ie} $\Tr
A_- B_-$ is in general different from zero. In particular $R^D_{\ominus}$
is the dual of $R^D_+$ with respect the symplectic trace, and
$R^D_{\oplus}$ is the dual of $R^D_-$.

Now we can define our 1-forms as parametrized by the dual space of
$R^D_{m-1}$ under the symplectic trace
(recall that these symbols parametrized the vector fields on $M^m_D$ ).
Therefore, 1-forms are parametrized
by elements in $R^D_{[-D-m+1,\infty )}$. If $A\in R^D_{m-1}$
and $X\in R^D_{[-D-m+1,\infty )}$, the action of the 1-form $X$
on $\partial_A$ is given by
$$X (\partial_A )=\Tr AX.\()$$

This let us define the gradient of a function by
$$dF(\partial_A)=\partial_A F.\()$$

In analogy with the finite dimensional case,
we should provide a map from one forms to vector fields in order to define
the Poisson brackets. The required
map is given by a suitable generalization of the standard
Adler map \[Adler], which reads in ``components''

$$J(X)=\{ \Lambda,X\}_{\oplus}\Lambda
 -\{ \Lambda,(\Lambda X)_+\}.\(classadler)$$

First notice that $J(X)\in R^D_{m-1}$, so it parametrizes a
vector field in $T_{\Lambda}M^m_D$. This is easily shown
if we write \(classadler) as

$$J(X)=-\{ \Lambda,X\}_{\ominus}\Lambda
 +\{ \Lambda,(\Lambda X)_-\}.\()$$

Also notice that because of the differnt splitting there is not
a natural reduction to $\Lambda_- =0$, in contrast to the standard
case\footnote\ddag{Of course, this does not necessarily imply
that there is no possible reduction where only a finite subset
of the $u_i$'s are different from zero. It only means that if
such reduction exists it will be more involved than in the usual
case.}.

Let $\Omega$ denote the map $X\mapsto\partial_{J(X)}$ from
1-forms to vectors fields induced by \(classadler). In analogy
with the finite dimensional case, it is convenient to introduce
the symplectic form $\omega$ defined, on Im$\Omega$, by
$$\omega (\Omega (X ),\Omega (Y))=
\Tr XY.\()$$

Notice that, in contrast with the usual case in classical mechanics,
this 2-form is not defined for all vector fields since, in general,
the map $J$ will not be an isomorphism.
It follows from the definition of $\omega$ that the Poisson brackets
will be given by
$$\pb{F}{G}_{GD}=\omega(\Omega(dF),\Omega(dG))=\Tr J(dF)dG,\(GD)$$
where we have introduced the suffix $GD$ (for Gel'fand and Dickey)
in order to avoid confusion with the canonical Poisson brackets
in a finite-dimensional phase space used for the
definition of the generalized Adler map.

It is now simple to check that this bracket is indeed antisymmetric.
Explicitly,
$$\eqalign{\pb{F}{G}_{GD}=& \Tr J(dF)dG\cr
=& \Tr\left(\pb{\Lambda}{dF}_{\oplus}\Lambda dG
-\pb{\Lambda}{(\Lambda dF)_+} dG\right)\cr
=& \Tr\left(-\pb{\Lambda}{(\Lambda dG)_-}dF
+\pb{\Lambda}{dG}_{\ominus} dF\right)\cr
=& -\Tr J(dG)dF =-\pb{G}{F}_{GD}\thinspace ,\cr}\()$$
where we have used $\Tr A_+B_{\oplus}=\Tr A_-B_{\ominus}=0$ for all
A and B.

By analogy with the finite-dimensional case, we define $d\omega$ by
$$\eqalign{d\omega (\partial_{J(X)},\partial_{J(Y)},\partial_{J(Z)})
=& \partial_{J(X)}\omega(\partial_{J(Y)},\partial_{J(Z)})\cr
& -\omega (\comm{\partial_{J(X)}}{\partial_{J(Y)}}
,\partial_{J(Z)})+ {\rm c.p.}\cr}\(dw)$$
where c.p. is  shorthand for cyclic permutations.
But notice that the last term in \(dw) is not
well defined unless Im$\Omega$ forms a subalgebra of the vector
fields. In fact, the proof of the following Lemma is routine.

\Lem<funnycom>{For any $X$ and $Y\in R^D_{[-D-m+1,\infty )}$
$$\comm{\partial_{J(X)}}{\partial_{J(Y)}}=
\partial_{J(\dlb{X}{Y})},$$
where
$$\eqalign{\dlb{X}{Y}=&\partial_{J(X)}Y -\partial_{J(Y)}X +\cr
&+\pb{X}{\Lambda}_{\ominus}Y+\pb{(\Lambda X)_-}{Y}+
\pb{Y}{\Lambda}_{\oplus}X +\pb{(\Lambda Y)_+}{X}\cr}$$
modulo the kernel of $J$.}

It is easy to
show \[DickeyIII] that closedness of $\omega$, {\ie} $d\omega =0$, is
equivalent to Jacobi identities for the bracket defined by \(GD).
Now we can state the main result of this paper.

\Thm<Main>{For any three vector fields $\partial_{J(X)}$,
$\partial_{J(Y)}$, and $\partial_{J(Z)}$ in Im$\Omega$
$$d\omega(\partial_{J(X)},\partial_{J(Y)},\partial_{J(Z)})
=0,$$
{\ie} $\omega$ is a closed 2-form.}

The proof of this theorem is given by a long, straightforward and
explicit computation of $d\omega$ which we omit.

\subsection {Bihamiltonian structure.}

We can obtain another Poisson structure by deforming the previous
one by $\Lambda\to\Lambda +\lambda$, for $\lambda$ some
constant parameter. Then
$$J(X)=J^2(X)+\lambda J^1(X),\()$$
where $J^2$ is given by \(classadler) and $J^1$ is given by
$$J^1(X)=-\pb{\Lambda}{X}_{\ominus} +\pb{\Lambda}{X_-}.\()$$

As usual the two kinds of
Poisson brackets so obtained are coordinated. For ``perverse''
historical reasons the stucture obtained by the deformation is
called the first hamiltonian structure while the one given
by \(classadler) is called the second.

We will finish this section by giving a convinient prescription
for computing the fundamental Poisson brackets among the
$u_j(x,\theta )$. Although the ``coordinates'' $u_j$ are not
functions according to the definition we are using, we can still make
sense of their Poisson bracket.

First notice that both structures are linear in $X$.
This implies that $J(X)$
is necessarily of the form
$$J^{1,2}(X)=\sum_{i,j\geq 1} (J^{1,2}_{ij}\cdot X_j)
|\xi |^{m-j},\(jotas)$$
where $\sum_{j\geq 1}X_j(x,\theta )|\xi |^{j-m-D}$,
and the $J_{ij}$'s are certain differential operators
acting on the $X_j$'s.

We now chose two linear functionals of the form
$$l_A =\Tr A\Lambda\ \ {\rm and}\ \ l_B=\Tr B\Lambda,\()$$
with $A=a(x,\theta )|\xi |^{i-m-D}$ and $B=b(x,\theta )|\xi |^
{j-m-D}$. Their gradients are given by
$$dl_A= a |\xi |^{i-m-D}\ \ {\rm and }\ \ dl_B=b |\xi |^{j-m-D},\()$$
which implies
$$\pb{l_A}{l_B}_{GD}^{1,2}=\int (J^{1,2}_{ij}\cdot a)b.\()$$

It is obvious that we would have obtained the same result
if we had declared our fundamental Poisson brackets among
the $u$'s to be
$$\pb{u_i(x,\theta )}{u_j(x',\theta ')}^{1,2}=
-J_{ij}^{1,2}\cdot \delta^D (x-x')\delta (\theta -\theta '),\(funda)$$
where $\delta (\theta -\theta ')$ is the ``delta-function''
associated with the standard measure in $S^{D-1}$.
Of course \(funda) is only true because $\int$ is a map which,
by assumption, is blind to boundary terms so we can freely
integrate by parts.

\section{D=1 revisited.}

In this section we will obtain the Poisson algebras induced by
\(classadler) in one dimension and associated with the Lax operator
$$\Lambda =\xi+\sum_{j\geq 0}u_j\xi^{-j}.\()$$

Notice that they will differ from
the standard $w_{KP}$\[Class] because of the different splitting.
But, interestingly enough,  the
new algebra, denoted by $w_{KZ}$, is nothing but the limit when
$n\rightarrow\infty$ of the standard $w_n$ after making the
the reduction of setting $u_0=0$.

A straightforward computation yields
$$\eqalign{J_{00}=& \partial ,\cr
J_{j0}=& u_{j-1}\partial ,\cr
J_{0j}=& \partial u_{j-1}.\cr}\()$$

And if $i,j\geq 0$, then
$$\eqalign {J_{ij}=& iu_{i+j-1}\partial +j\partial u_{i+j-1}
+(1-j)u_{i-1}\partial u_{j-1}\cr
&+\sum_{k=0}^{j-2}\left( (i-k-1)u_{i+j-k-2}\partial u_k
+(j-k-1) u_k\partial u_{i+j-k-2}\right).\cr}\()$$

Imposing now the constraint $u_0=0$, the associated Dirac brackets
give
$$\eqalign {J^{(0)}_{ij}=& iu_{i+j-1}\partial +j\partial u_{i+j-1}
-ju_{i-1}\partial u_{j-1}\cr
&+\sum_{k=1}^{j-2}\left( (i-k-1)u_{i+j-k-2}\partial u_k
+(j-k-1) u_k\partial u_{i+j-k-2}\right),\cr}\(wkp)$$

The explicit form of the classical Gel'fand--Dickey algebras
with the standard splitting is given in \[ClassLim]. It reads
$$\veqnalign{J_{n-1,n-1} &= -n\d~,\cr
J_{i,n-1} &= -(i+1)u_{i+1}\d~,\cr
J_{n-1,j} &= -(j+1)\d u_{j+1}~,\cr
J_{ij} &= (n-j-1) \d u_{i+2+j-n} + (n-i-1) u_{i+2+j-n}\d\(gdn)\cr
&\quad{}+ \sum_{l=j+2}^{n-1} \left[ (l-i-1) u_{i+j+2-l}\d u_l +
(l-j-1) u_l \d u_{i+j+2-l}\right]\cr
&\quad{}-(i+1)u_{i+1}\d u_{j+1}~,\cr}$$
where $i,j=0,1,\ldots,n-2$ and with the proviso that $u_{l< 0}=0$ in
the above formulas.

If we now do the following field redefinition $u_{n-j-1}\mapsto u_j$.
We obtain
$$\eqalign{J_{00}=& -n\partial ,\cr
J_{j0}=& (j-n)u_{j-1}\partial ,\cr
J_{0j}=& (j-n)\partial u_{j-1}.\cr}\()$$

And if $i,j\geq 0$, then
$$\eqalign {J_{ij}=& iu_{i+j-1}\partial +j\partial u_{i+j-1}
-(n-i)u_{i-1}\partial u_{j-1}\cr
&\sum_{k=0}^{j-2}\left( (i-k-1)u_{i+j-k-2}\partial u_k
+(j-k-1) u_k\partial u_{i+j-k-2}\right).\cr}\()$$

If we again impose the constraint $u_0=0$, we have
$$J_{ij}^{(0)}=J_{ij}+{(j-n)(i-n)\over n}u_{i-1}\partial u_{j-1}.\()$$

And now we can take the limit $n\rightarrow\infty$ in the expression
above and recover \(wkp). Therefore
$$w_{KZ}\simeq w_{n\rightarrow\infty}.\()$$

Let us now focus our attention on the first structure. After the
natural reduction of setting $u_0=u_1=0$, the Adler map simply
becomes
$$J(X)=-\pb{\Lambda}{X_+}_{\ominus}.\()$$

{}From this we obtain
$$J_{ij}=(i-1)u_{i+j-2}\partial +(j-1)\partial u_{i+j-2}\()$$

and this is nothing but the $w_{\infty}$ algebra,
as the reader can check in \[Winfty].

All of this seems to indicate that the new splitting is quite
``natural'' in one dimension.

\section{D=2}

Now we are going to use the tools developed in Section 2 in order
to compute an explicit example of
the new higher dimensional classical $W$-algebras.
We will limit ourselves to the two-dimensional case because,
as our reader will see in what follows, explicit expressions
soon become very cumbersome.

Let us first compute the Poisson brackets associated with the
second hamiltonian structure in $M^m_D$.
The fundamental Poisson brackets between the $u$'s, after imposing
the constraint $u_1(x,\theta )=u_2(x,\theta )=0$,
are given by
$$\pb{u_i^t(z,\bar z)}{u_j^p(z',\bar z')}=-J_{ij}^{tp}\cdot
\delta (z-z')\delta (\bar z-\bar z'),\(funpb)$$
where we have introduced complex coordinates $z={1\over 2}(x^2+ix^1)$,
and its complex conjugate $\bar z$, and where
$$u_j(x,\theta )=\sum_{p\in\integ}u_j^p{\rm e}^{ip\theta },$$
with
$$\eqalign{J_{ij}^{tp}&=
(j+p-2)\partial_{\bar z}u_{i+j-3}^{t+p-1}+
(i+t-2)u_{i+j-3}^{t+p-1}\partial_{\bar z}\cr
&+(j-p-2)\partial_{z}u_{i+j-3}^{t+p+1}+
(i-t-2)u_{i+j-3}^{t+p+1}\partial_{z}\cr
-\sum_{s\in\integ}\sum_{k\geq j}&\Bigl(
(i+t-k-s-2)u_{i+j-k-3}^{t+p-s-1}\partial_{\bar z}u_k^s\cr
&+(j+p-k-s-2)u_k^s\partial_{\bar z}u_{i+j-k-3}^{t+p-s-1}\cr
&(i-t-k+s-2)u_{i+j-k-3}^{t+p-s+1}\partial_{z}u_k^s\cr
&+(j-p-k+s-2)u_k^s\partial_{z}u_{i+j-k-3}^{t+p-s+1}\Bigr)\cr
+\sum_{s\in\integ}\sum_{j-2\leq k< j}&\Bigl(
(2m+k+s-t-p-i-j-4)u_{i+j-k-3}^{t+p-s-1}\partial_{\bar z}u_k^s\cr
&+(2m+k-s+t+p-i-j-4)u_{i+j-k-3}^{t+p-s+1}\partial_{z}u_k^s\cr
&+(k+s-p-j+2)u_k^s(\partial_{\bar z}u_{i+j-k-3}^{t+p-s-1})\cr
&+(k-s+p-j+2)u_k^s(\partial_{z}u_{i+j-k-3}^{t+p-s+1})\Bigr),\cr
}\(chori)$$
and the proviso that $u_{j<3}^t=0$ for all $t\in\integ$.
We have chosen to complexify
the algebra, {\ie} consider complex $u$'s, because expressions
become more transparent. Of course a real section can be taken if
desired.

Now it is simple to check that there is a finite-dimensional subalgebra
generated by $u_3^1$ and $u_3^{-1}$. If we use the suggestive notation
$u_3^1\equiv 2P_{\bar z}$ and $u_3^{-1}\equiv 2P_z$, we obtain
$$\eqalign{
\pb{P_{\bar z}(z,\bar z)}{P_{\bar z}(z',\bar z')}
=-&\bigl(\partial_{\bar z}P_{\bar z}(z,\bar z)
+P_{\bar z}(z,\bar z)\partial_{\bar z}\bigr)\cdot\delta (z-z')
\delta (\bar z-\bar z')\cr
\pb{P_{z}(z,\bar z)}{P_{z}(z',\bar z')}
=-&\bigl(\partial_{z}P_{z}(z,\bar z)
+P_{z}(z,\bar z)\partial_{z}\bigr)\cdot\delta (z-z')
\delta (\bar z-\bar z')\cr
\pb{P_{\bar z}(z,\bar z)}{P_{z}(z',\bar z')}
=-&\bigl(\partial_{z}P_{\bar z}(z,\bar z)
+P_{z}(z,\bar z)\partial_{\bar z}\bigr)\cdot\delta (z-z')
\delta (\bar z-\bar z').\cr
}$$

This is nothing but the algebra of local diffeomorphims of
a two dimensional manifold in complex coordinates.

By inspecting expression \(chori), the reader can check that the
$u_j^t$ fall in two different representations of the diffeomorphism
algebra. The sets composed by $u_j^{j-2},u_j^{j-4},....,u_j^{-j+2}$
form finite dimensional-representation, which correspond to
$(j-2)$-symmetric covariant tensor 1-densities. The others
fall in two infinite-dimensional representations, corresponding
to an odd or even upper index. It would be very interesting to
find a reduction which will leave us only with the finite-dimensional
representations, but for the time being we have not
been able to find it.

In what concerns the first structure, we will only say that
after the natural reduction of setting all the $u_j(x,\theta )$
with $j\leq m+2$ to zero, the algebra obtained is isomorphic to
the linear part of \(chori). In this case, it is worth noticing
that there is a subalgebra spanned by the finite-dimensional
representations of the diffeomorphism
algebra\footnote\dag{ This subalgebra is closely related to
the symmetric Schouten bracket. We hope to come back to
this issue in a future publication.}.

\section{Associated integrable hierarchies.}

In this section we will show that these new Poisson structures give
a bihamiltonian formulation of higher-dimensional KZ-hierarchies.

Let us first recall the standard one-dimensional formulation of
these hierarchies. The KP-hierarchy$^{\[KP]}$ can be defined as the Lax-type
evolution equations given by
$$\pder{\Lambda}{t_n}=\comm{\Lambda_+^n}{\Lambda} =
\comm{\Lambda}{\Lambda_-^n},\(KP)$$
where $\Lambda=\xi +\sum_{i\geq 0}u_i\xi^{-i}$, and the $+$ and $-$
stand for the standard projections over the differential and
``integral'' parts. The KZ-hierarchy$^{\[TakaTake] ,\[KoGi]}$
is nothing but the classical limit of \(KP), so it reads
$$\pder{\Lambda}{t_n}=\pb{\Lambda_+^n}{\Lambda} =
\pb{\Lambda}{\Lambda_-^n}.\(KZ)$$

{}From the point of view of the differential equations, the classical
limit is such that higher derivative terms are disregarded
or, equivalently,
fields are taking to be slowly varying in their spatial coordinate.

It is now a simple exercise to show that all these flows
for different $n$ commute.

\Prop<FLOWSCOM>{For all $i,j\in\nats$,
$$\dpder{\Lambda}{t_i}{t_j} = \dpder{\Lambda}{t_j}{t_i}\ .\()$$}

\Pf
$$\veqnalign{\dpder{\Lambda}{t_i}{t_j} &= \pder{}{t_i}
\pb{\Lambda^{j}_+}{\Lambda}\cr
&= \pb{\left(\pder{\Lambda^{j}}{t_i}\right)_+}{\Lambda} +
\pb{\Lambda^{j}_+}{\pb{\Lambda^{i}_+}{\Lambda}}\cr
&= \pb{\pb{\Lambda^{i}_+}{\Lambda^{j}}_+}{\Lambda} +
\pb{\Lambda^{j}_+}{\pb{\Lambda^{i}_+}{\Lambda}}&\cr
&= \pb{\pb{\Lambda^{i}_+}{\Lambda^{j}}_+}{\Lambda} +
\pb{\Lambda^{i}_+}{\pb{\Lambda^{j}_+}{\Lambda}} +
\pb{\pb{\Lambda^{j}_+}{\Lambda^{i}_+}}{\Lambda}\cr
&= \pb{\pb{\Lambda^{i}_+}{\Lambda^{j}_-}_+}{\Lambda} +
\pb{\Lambda^{i}_+}{\pb{\Lambda^{j}_+}{\Lambda}}\cr
&= \pb{\pb{\Lambda^{i}}{\Lambda^{j}_-}_+}{\Lambda} +
\pb{\Lambda^{i}_+}{\pb{\Lambda^{j}_+}{\Lambda}}\cr
&= \pb{\pb{\Lambda^{j}_+}{\Lambda^{i}}_+}{\Lambda} +
\pb{\Lambda^{i}_+}{\pb{\Lambda^{j}_+}{\Lambda}}\cr
&=\dpder{\Lambda}{t_j}{t_i}\ .&\qed\cr}$$

This is an important result because if these flows are hamiltonian,
\<FLOWSCOM> implies that there are an infinite number of commuting
conserved charges, thus proving formal Liouville integrability
of the system. In the case of the KZ-hierarchy, these flows are
bihamiltonian with respect to $w_{KP}$ and $w_{1+\infty}$ \[Class].

As the reader has probably already noticed, the key property in the
proof of \<FLOWSCOM> was that there is an invariant splitting with
respect to Poisson brackets. But in section 2 it has
already been shown that such splitting exist in arbitrary dimension
if the definitions are slightly modified. This implies that \(KZ)
defines a Liouville integrable hierarchy in higher dimensions
whenever $\Lambda$   belongs to $M_D^1$
and the $+$ part stands for the projection
over $R^D_{[1,\infty )}$. Then the proof of \<FLOWSCOM> goes
step-by-step to the higher-dimensional case.

It is now trivial to show that
$$\pder{\Lambda}{t_k}=J^1(dH_{k+1})=J^2(dH_k)\(FLUJOS)$$
where $H_k={1\over k}\Tr\Lambda^k$, and its gradient is given by
$$dH_k =\Lambda^{k-1}\ {\rm mod}R^D_{-D}.\()$$

The $H_k$ are clearly conserved charges for any of the KZ-flows,
that is
$$\pder{H_k}{t_j}=\Tr\pb{\Lambda^j_+}{H_k}=0\()$$
because the symplectic trace annihilates Poisson brackets. The proofs
of nontriviality and independence of these conserved charges are
identical to the usual ones, so we will not repeat them here.

Standard procedures also yield
$${\pb{H_k}{H_j}}^1_{GD}={\pb{H_k}{H_j}}^2_{GD}=0,\(COMMU)$$
for all $k,j\in\nats$.

\section{Conclusions.}

We have shown how to construct classical $W$-algebras in higher
dimensions. The standard one dimensional formalism has been shown
to extend to
higher dimensions with minor modifications and the use of some
new tools such as the symplectic trace.

Of course much remains to be done. We have not proved that
the connection between these new classical $W$-algebras and the
algebra of generators of local diffeomorphisms goes beyond
dimension two, although we firmly believe that it is so.
Neither have we studied potentially interesting reductions of these
algebras, in particular if we can restrict ourselves to fields
falling in finite dimensional representations of the
diffeomorphism algebra.
It would also be an interesting problem to see the connection
between $w_{KZ}$ and the $c=1$ matrix model formulation of
matrix models. The fact that $w_{KZ}\simeq w_{n\to\infty}$
seems to strongly indicate that such a connection indeed exists.
It would also be interesting to construct Lagrangian field
theories with these new algebras as their algebra of symmetries,
and last but not least, to ``quantize'' them.

We would not like to finish this paper without a word about why
we have restricted ourselves to the classical case.

\subsection{A Comment on Deformations.}

It would look very natural to try to extend the present formalism
to the ring of {\pdo}'s in arbitrary dimension without restricting
ourselves to the classical limit. There is a noncommutative
generalization of the symplectic trace due to Wodzicki$^{\[Wod]}$,
so we could simply try to substitute our Poisson brackets in
the underlying $2D$-dimensional symplectic space for commutators.
Unfortunately, without such a
restriction we rapidly get into trouble, the main problem being
that there is no invariant splitting for $D>1$
with respect to the commutator
of {\pdo}'s, {\ie} it is not possible to defined
a $+$  and $-$ subalgebras closed under commutation and such
that $R^D =R_+^D\oplus R_-^D$.
The interested reader can check
that this is a required ingredient in the proofs of
\<funnycom> and \<Main>.

Nevertheless, it is interesting to point out that the hierarchy
defined by \(KP) for a {\pdo} $\Lambda\in M_D^1$
still posseses an infinite number of conserved
charges, which are given by the Wodciki-Trace$^{\[Wod]}$ of the
integer powers of $\Lambda$.
But in this case it is simple to prove that the flows do not
commute, as this would necesitate the existence
of an invariant splitting. The reader is invited to verify
this in the proof of
\<FLOWSCOM>. This implies that the hierarchy so defined
would not be integrable, at least in the sense of Liouville.

The above discussion suggests that deformations of these
structures is a ``tricky'' business in $D>1$.
\vskip 1.5cm
{\bf Acknowledgments}
\vskip 0.3cm
We would like to thank J. Figueroa-O'Farrill and J. Mas for
many useful conversations. We are also thankful to
Anne Petrov for a careful reading of the manuscript.
E.R. would also like to thank
the hospitality of the Dept. of Particulas Elementales of
Santiago, where part of this work was completed.

\refsout
\bye